\documentclass[reprint, superscriptaddress, amsmath,amssymb, aps, pra, longbibliography]{revtex4-1}

\usepackage{soul}
\usepackage{amsmath}   
\usepackage{amssymb}
\usepackage{mathtools}
\usepackage{graphicx}
\usepackage{dcolumn}
\usepackage{bm}
\usepackage{amsmath}
\usepackage{graphicx}
\usepackage{amsfonts}
\usepackage{subfigure}
\usepackage{graphicx}
\usepackage{array}
\usepackage{float}
\usepackage{color}
\usepackage{multirow}
\usepackage{amssymb}%
\usepackage[colorlinks=true,linkcolor=blue]{hyperref}%
\hypersetup{allcolors=blue}
\usepackage[normalem]{ulem}
\usepackage{xcolor}

\newcommand{\ket}[1]{\ensuremath{\left\vert #1 \right\rangle}}
\usepackage{mathtools}
\DeclarePairedDelimiterX\braket[2]{\langle}{\rangle}{#1 \delimsize\vert #2}

\begin{document}
\title{Filter circuit for suppression of electric-field noise in Rydberg-atom experiments}
\date{\today }

\author{Xinyan Xiang}
\email{xinyann@umich.edu}
\affiliation{Department of Physics, University of Michigan, Ann Arbor, MI 48109, USA}
\author{Shuaijie Li}
\affiliation{Department of Physics, University of Michigan, Ann Arbor, MI 48109, USA}
\author{Alisher Duspayev}
    \thanks{Present address: Department of Physics and Joint Quantum Institute, University of Maryland, College Park, MD 20742, USA.}
    \affiliation{Department of Physics, University of Michigan, Ann Arbor, MI 48109, USA}   
\author{Ian Hoffman}
\thanks{Present address: Department of Physics, University of Alabama, Tuscaloosa, AL 35401, USA}
\affiliation{Department of Physics, University of Michigan, Ann Arbor, MI 48109, USA}
\author{Lefeng Zhou}
\affiliation{Department of Physics, University of Michigan, Ann Arbor, MI 48109, USA}
\author{Bineet Dash}
\affiliation{Department of Physics, University of Michigan, Ann Arbor, MI 48109, USA}
\author{Carlos Owens}
\affiliation{Department of Physics, University of Michigan, Ann Arbor, MI 48109, USA}
\author{Anisa Tapper}
\affiliation{Department of Physics, University of Michigan, Ann Arbor, MI 48109, USA}
\author{Georg Raithel}
\affiliation{Department of Physics, University of Michigan, Ann Arbor, MI 48109, USA}

\begin{abstract}
Rydberg atoms are widely employed in precision spectroscopy and quantum information science. To minimize atomic decoherence caused by dc Stark effect, the electric field noise at the Rydberg atom location should be kept below $\sim 10$~mV/cm. Here we present a simple yet effective electronic circuit, referred to as a clamp switch, that allows one to realize such conditions. The clamp switch enables precise low-noise electric field control while allowing application of fast high-voltage ionization pulses through the same electrode(s), enabling atom detection via electric-field ionization and electron or ion counting. We outline the circuit design and analyze its noise suppression performance for both small and large input signals. In application examples, we employ the clamp switch to reduce the spectral width and increase the signal strength of a Rydberg line by a factor of two, to estimate the electric-field noise in the testing chamber, and to perform electric-field calibration using Rydberg Stark spectroscopy. The clamp switch improves coherence times and spectroscopic resolution in fundamental and applied quantum science research with Rydberg atoms. 
\end{abstract}

\maketitle

\section{Introduction}
\label{sec:intro}

Atoms in highly excited (Rydberg) states~\cite{gallagher} are widely used in modern research in atomic, molecular and optical physics, corresponding applications in fundamental quantum science~\cite{Dunning_2016,browaeysreview} and emerging atom-based quantum technologies~\cite{saffmanrmp2010, Adams_2020, morgado2021}. Excitation and manipulation of Rydberg atoms using optical and microwave fields have become ubiquitous in both ultra-high-vacuum setups and room-temperature vapor cells. High-fidelity detection of Rydberg excitations is often important in such experiments~\cite{singer2004,sylvain2018,levineprl2018}. All-optical, non-destructive methods, such as Rydberg electromagnetically-induced transparency (EIT)~\cite{Fleischhauer2005,mohapatra2007}, have proven useful in Rydberg-atom spectroscopy, and helped advance directions such as Rydberg-atom-based electric-field sensing~\cite{Sedlacek2012, Holloway2014, fancher2021} and quantum optics with cold atoms~\cite{chang, Firstenberg_2016, Kumlin_2023}. However, EIT is currently not practical in experiments that require state-selective detection of individual Rydberg atoms in high-vacuum setups. This includes instances in high-precision spectroscopy of Rydberg transitions using microwaves~\cite{Sassmannshausen2013, leepra2016, Bai2023} or ponderomotive interactions \cite{Cardman2023}, research on Rydberg molecules~\cite{shafferreview, Bai2024}, and fundamental and applied research involving circular Rydberg atoms~\cite{anderson2013, zhelyazkova2016, Ramos2017}. The latter may serve as a resource in quantum information science~\cite{Signoles2017, holzl2024}. A common and well-established method in such experiments is electric field ionization (FI) and state-selective FI detection (SSFI)~\cite{gallagher}. In our paper, FI refers to both of these detection schemes.     

FI is based on the use of an external electric field 
to ionize atoms via tunneling or related dynamics of the Rydberg electron~\cite{littman1978, gallagher}.
The method is usually employed in ultra-high-vacuum setups that incorporate electric-field shielding and control electrodes for Rydberg-atom preparation and manipulation~\cite{grimmel2017}. 
The same electrodes may be used to apply a large electric field for FI of Rydberg atoms for readout via electron or ion detection with a charged-particle detector~\cite{JAGUTZKI2002244}. In practice, this means that the electric potential applied to at least one electrode must be switched between a well-controlled low voltage that has low noise, applied at times when the scientifically relevant Rydberg-atom physics occurs (Rydberg excitation, control, manipulation etc.), and a high voltage for FI and subsequent electron or ion detection with the charged-particle detector. The electric polarizabilities of Rydberg levels typically scale as $n^7$, with principal quantum number $n$, and can exceed $h \times 10^4$~MHz/(V/cm)$^2$.
As a result, during the science phase of the experiment the electric field noise has to be kept below several mV/cm ~\cite{MARCASSA201447,veit2021}.
The need for electrode(s) carrying low-noise Rydberg-atom control potentials at certain times and potentials of hundreds of volts for FI at other times presents a challenge due to the considerable voltage noise of most high-voltage sources and amplifiers. Notably, in most Rydberg-atom experiments voltage noise during FI can be several orders of magnitude higher than during the science phase without affecting performance.

In this paper, we describe an electronic circuit that solves this problem. The circuit, which we refer to as clamp switch, is employed together with additional circuitry that generates a low-voltage, low-noise signal for electric-field compensation and fine control during Rydberg-atom excitation and manipulation, a high-voltage (HV) signal for Rydberg-atom FI and detection, and a HV multiplexer that switches between these two voltages. The HV components are intrinsically noisy, often carrying noise in the range of hundreds of mV. The clamp switch is inserted between the HV multiplexer and the FI electrodes in the setup to reduce the transmission of noise from the HV line into the apparatus during the science phase of the experimental sequence. During this phase, an additional low-voltage \text{offset} input on the clamp switch allows for accurate, low-noise and dynamical control of the electric field applied to the Rydberg atoms. The clamp switch further transmits fast HV pulses in a practically attenuation-free manner, enabling FI detection of the atoms. 
The HV pulses are sufficiently isolated from the low-voltage offset input for protection of the utilized low-voltage source.
In the following, we describe the circuit, quantify its small-signal noise suppression and large-signal switching characteristics over the relevant time scales, and present spectroscopic data from a cold Rydberg-atom setup that exemplify the benefits of using the clamp switch. 

\section{Clamp Switch Operation}
\label{sec:operation}
\subsection{The circuit and principle of operation}
\label{subsec:circuit}

\begin{figure}[t]
    \centering
    \includegraphics[width=0.48\textwidth]{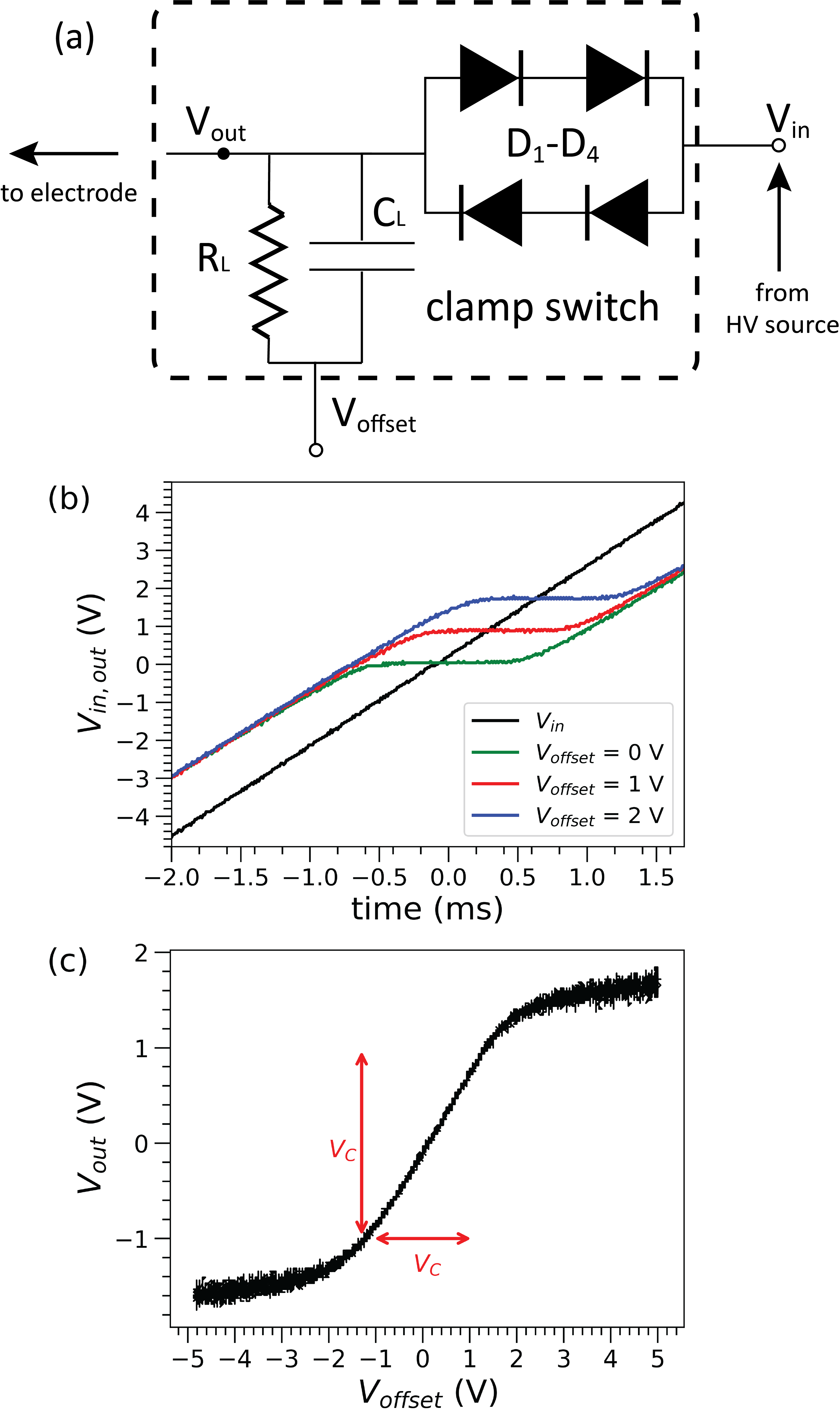}
    \caption{(a) Diagram of the clamp switch circuit. The clamped signal produced at the output location, $V_{out}$, is applied to in-vacuum electrodes in the experimental setup. The load capacitance, $C_L$, is mostly due to connection coax cables.
    (b) Sample oscilloscope traces of the clamp switch input voltage, $V_{in}(t)$ (black trace), and $V_{out}(t)$ for selected, fixed values of the $V_{offset}$ input (green, red, and blue traces). (c) Sample oscilloscope trace of $V_{out}(t)$ versus $V_{offset}(t)$ with fixed input $V_{in} \approx 0$~V. Within the clamping range $V_C$, indicated by the red arrows, the input noise of several 100~mV, visible in the wings of the curve, is largely blocked, while $V_{out}$ can be regulated via $V_{offset}(t)$ with near-unity gain, $dV_{out}/dV_{offset} \approx 1$.}
    \label{fig:circuitdiagram}
\end{figure}

The circuit diagram for the clamp switch is shown in Fig.~\ref{fig:circuitdiagram}~(a). It consists of four diodes (D1 through D4) and has two inputs and one output. The diodes have a voltage-dependent resistance $R_D$ and capacitance $C_D$. The HV input, $V_{in}$, is connected to the HV multiplexer's output [HV source in Fig.~\ref{fig:circuitdiagram}~(a)]. The HV signal is passed through the diodes into a load resistor, $R_L$.  The parallel load capacitance, $C_L$, mostly arises from coax cables connected to the Rydberg experiment [see Fig.~\ref{fig:circuitdiagram}~(a)]. The load resistor is connected to the low-impedance, low-noise offset input of the clamp switch, $V_{{offset}}$. The node between the diodes, $R_L$, and $C_L$ defines the clamp switch's output, $V_{out}$. It is connected to the electrode(s) within the Rydberg experiment. The electrode(s) typically contribute little to $C_L$ and have a practically infinite electric resistance against other components. 

The switch operates as follows. If the voltage difference between the HV input $V_{in}$ and the bias voltage $V_{offset}$ is smaller than the diodes' forward voltage, all diodes have an impedance $\gg R_L$. The low-noise input $V_{offset}$ is then coupled through $R_L$ to the setup's electrode(s), while isolating the noise from the comparatively noisy HV source. The output is then $V_{out} \approx V_{offset}$, and the noise on the input line $V_{in}$ is largely isolated from $V_{out}$. Further, during the science phase of the experiment $V_{offset}$ can be regulated over a clamping range of several volts, enabling high-fidelity Rydberg atom manipulation and control without adverse effects from noise on the input line $V_{in}$. 

For FI, the input line $V_{in}$ is ramped up to HV using a reasonably fast HV multiplexer, a HV-amplified waveform, or similar. In this case, the voltage difference between the HV input and $V_{offset}$ is much larger than the diodes' forward voltage, typically, by several orders of magnitude. Most of the voltage drop then occurs at the load resistor $R_L$, and the HV on $V_{in}$ is passed through to $V_{out}$ and to the internal electrode(s) of the setup. Notably, FI performance is not affected by moderate electric field noise. In this way, the clamp switch allows application of a low-noise, low-voltage signal for high-fidelity Rydberg atom manipulation and control, while maintaining the ability to apply a comparatively noisy HV signal for FI via shared electrode(s).    

As shown in Fig.~\ref{fig:circuitdiagram}~(a), there are two pairs of diodes oriented in opposite directions, allowing us to pass both positive and negative FI ramps on $V_{in}$ through to $V_{out}$. This provides flexibility in electrode geometry and enables both electron and positive-ion detection capability using the same electrode(s). 
Furthermore, increasing the number of diodes expands the range in $\vert V_{in} - V_{offset} \vert$ over which the clamp switch isolates the noise of the HV source from the experiment. 
Since the voltage across the diode section has an upper limit given by the number of diodes per direction times the single-diode forward voltage, reverse breakdown is of no concern. It should also be ensured that the time-averaged thermal power on the load resistor, which is $\lesssim \langle V_{in}^2(t) \rangle_{t} /R_L $, remains below the resistor's power rating.

The presented design is adaptable for a range of diode types, $R_L$-values, and capacitances.  Generally, for fast FI sequences, low-capacitance diodes with short reverse recovery times are preferred (see Sec.~\ref{subsec:acresponse}). Moreover, the $R_L$-value should be chosen as low as possible, subject to a lower limit set by the time-averaged thermal power on $R_L$, the current sourcing capability of the specific HV source connected to $V_{in}$, and over-voltage protection of the low-voltage source connected to $V_{offset}$. The load capacitance, $C_L$, and the diodes' junction capacitances have to be similar. As $C_L$ has a lower limit given by the minimum length of coax connection cables needed, certain applications may require an extra tuning capacitor, $C_P$, applied in parallel with the diode legs (see Appendix).

In the remainder of this paper, we use the circuit shown in Fig.~\ref{fig:circuitdiagram}~(a) with 1N4007 diodes for D1-D4, a load capacitance estimated to be 200~pF, and a load resistance $R_L$ typically on the order of several hundred k$\Omega$. Specifically, $R_L$ is set to $200$~k$\Omega$ in Sections~\ref{subsec:small} and~\ref{subsec:acresponse}, while a lower value of $100$~k$\Omega$ is considered in Section~\ref{sec:appl}. To demonstrate the operation of the clamp switch, in Fig.~\ref{fig:circuitdiagram}~(b) we show $V_{out}$ versus $V_{in}$ for several fixed values of $V_{offset}$ generated by a low-impedance DC power supply, while $V_{in}$ is linearly scanned in time. The HV source used in our testing has an impedance of $\sim 27~$k$\Omega$ and a current limit to safeguard against shorts in the Rydberg experiment. The plot range in Fig.~\ref{fig:circuitdiagram}~(b) is cropped to the relevant range, $\vert V_{in} \vert \lesssim 4$~V. Within a range $\vert V_{in} - V_{offset} \vert \lesssim 1.2$~V, the output $V_{out}$ passes through a plateau centered at $V_{offset}$. For example, for $V_{offset}=1$~V the output is clamped to $V_{out} \approx 1$~V over an input voltage range $-0.8$~V~$ \lesssim V_{in} \lesssim 2.2$~V. Within the clamping range, which has a full width $\lesssim 2.5$~V, the noise of the HV source connected to $V_{in}$ is largely isolated from $V_{out}$. As $V_{offset}$ is varied, the center of the clamping plateau in $V_{out}$ shifts to $V_{out} =  V_{in} = V_{offset}$, while the width of the clamping range, $V_C = 2 \vert V_{in} - V_{offset} \vert \lesssim 2.5$~V, does not change.

In Fig.~\ref{fig:circuitdiagram}~(c) we show $V_{out}$ versus $V_{offset}$ for fixed $V_{in} \approx 0$~V, with $V_{in}$ carrying several 100~mV of noise. It is seen that within the clamping range, $V_C$, the low-noise input $V_{offset}$ is passed through to $V_{out}$ while blocking the noise from $V_{in}$. This enables high-fidelity voltage control on $V_{out}$ over a limited range. If $\vert V_{in} - V_{offset} \vert \gtrsim V_C/2$, the value of $V_{out}$ plateaus at $V_{out} \approx V_{in} \pm V_C/2$, and the noise from $V_{in}$ is passed through to $V_{out}$. Although moderate noise is typically acceptable in FI of atoms, it would severely degrade the fidelity of Rydberg-atom control and manipulation sequences in the science phase of the experiment.

\subsection{Small-signal response of the clamp switch}
\label{subsec:small}

Noise on $V_{in}$ is characterized by the noise spectrum and the root mean square (RMS) deviation, $\delta V_{in}$. During the Rydberg-atom manipulation and control phase of the experiment, we may assume $|V_{in}-V_{offset}| < V_C/2$ and $\delta V_{in} \ll V_C/2$. Investigation of the circuit's response in this small-signal regime will allow us to assess the noise suppression characteristics. To this end, we measure the frequency dependence of the circuit's small-signal attenuation and phase shift. 

\begin{figure}[t]
 \centering
  \includegraphics[width=0.48\textwidth]{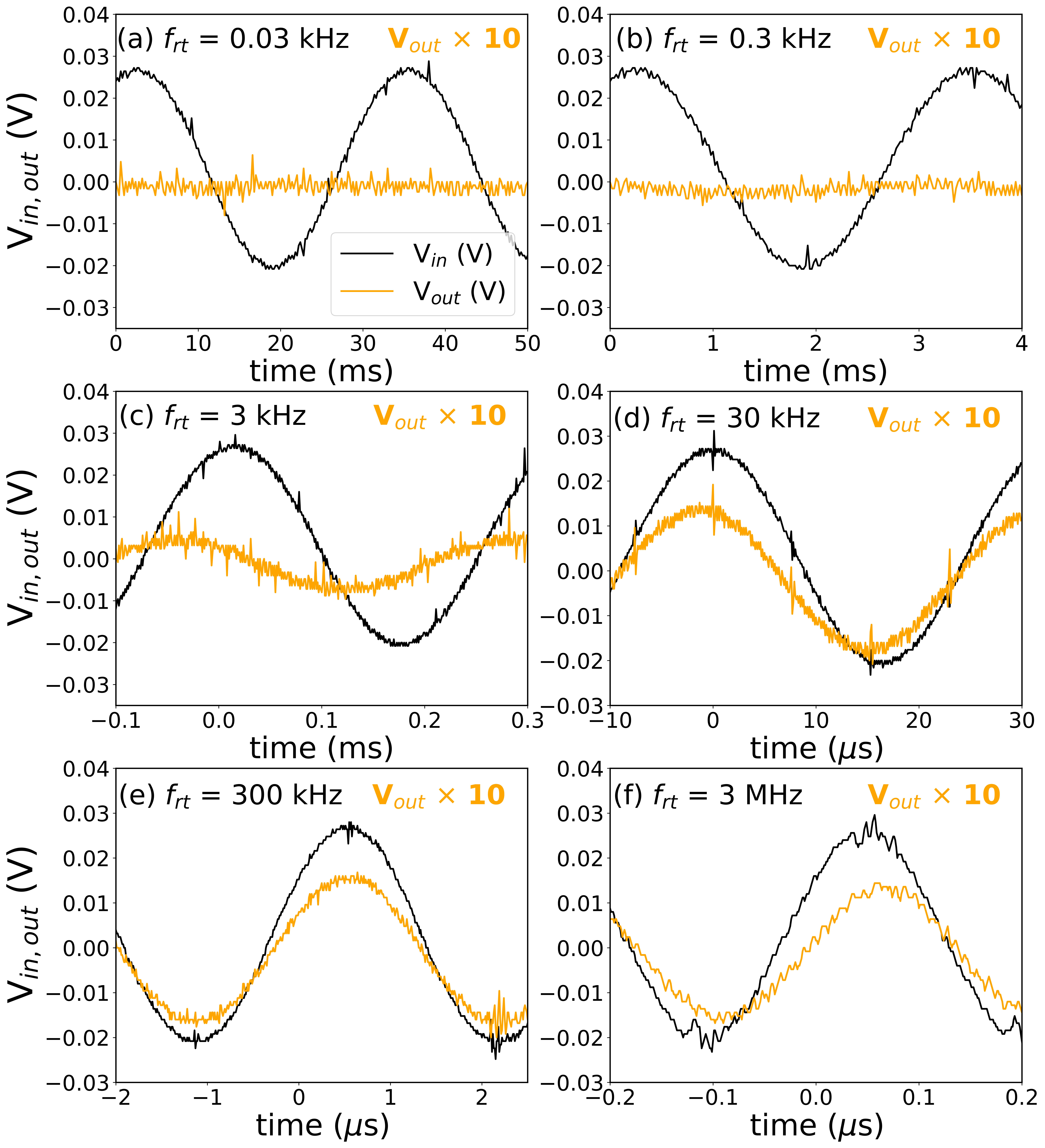}
  \caption{Small-signal response of the clamp switch at the indicated frequencies. The vertical axis shows both $V_{in}$ and $V_{out}$. For clarity, $V_{out}$ has been scaled up by a factor of ten. The clamping efficiency is observed to drop with increasing frequency. Also, there is a frequency-dependent phase shift. The phase shift is challenging to measure at low frequency, where the clamp switch efficiently isolates the input from the output. 
  } 
  \label{fig:sine response}
\end{figure}

In Fig.~\ref{fig:sine response}, we apply a small sine wave signal with a peak-to-peak amplitude of 25~mV and a variable frequency $f_{rt}$ ranging from 0.03 kHz to 3~MHz to $V_{in}$ (black lines), with $V_{offest}$ close to zero, and we measure the output $V_{out}$ (orange lines). As seen in Fig.~\ref{fig:sine response}, at low frequencies, $f_{rt} \lesssim $~3~kHz, the amplitude of $V_{out}$ is highly suppressed, exhibiting the desired clamping effect. At higher frequencies, $f_{rt} \gtrsim 3$~kHz, the output $V_{out}$ is phase-shifted relative to $V_{in}$ and remains suppressed in amplitude by a factor of about 15. 

\begin{figure}[t]
 \centering
  \includegraphics[width=0.45\textwidth]{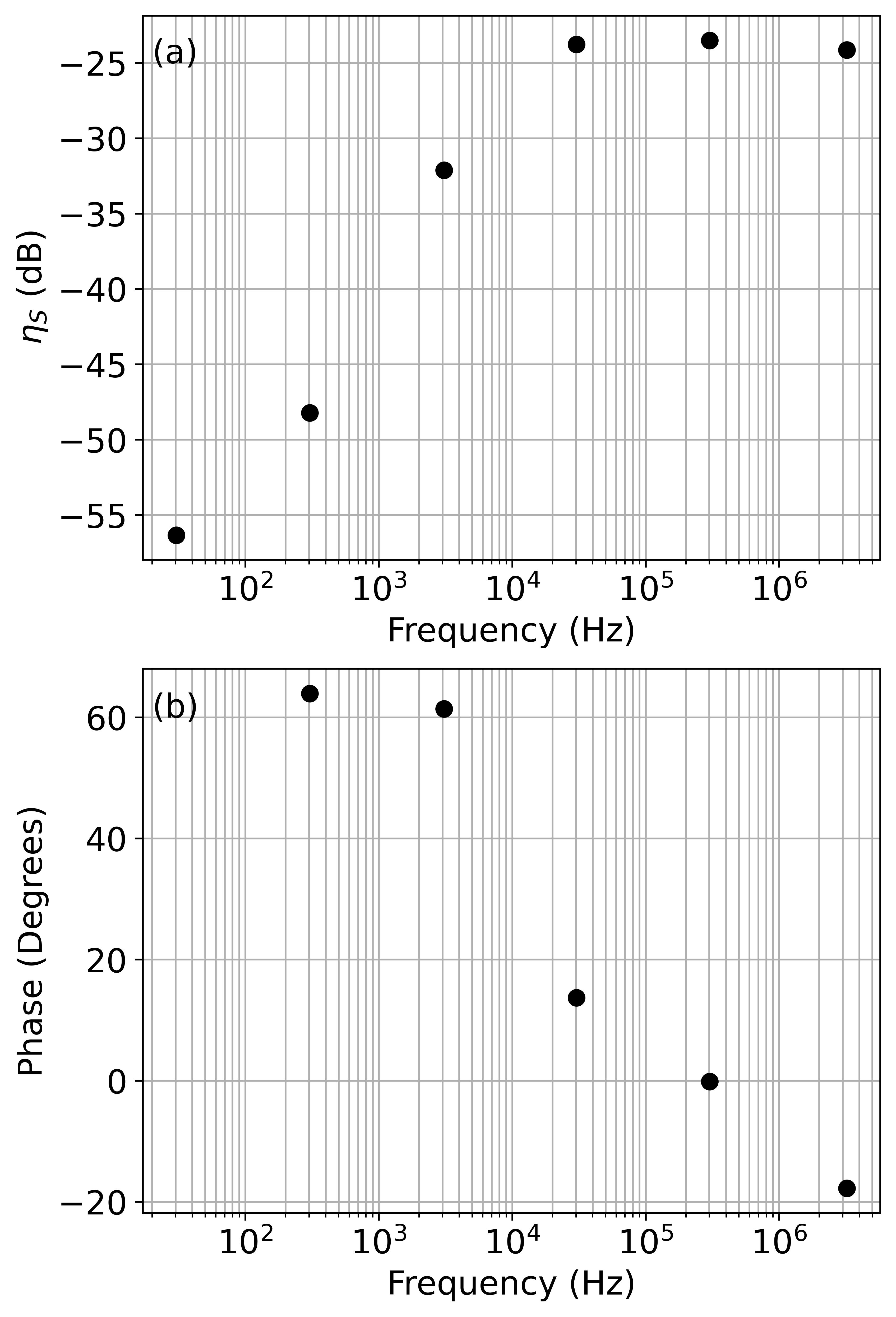}
  \caption{Bode plot of the clamp switch. In (a) we show the attenuation factor, $\eta_S$, and in (b) the phase shift as a function of frequency, measured with a small sinusoidal signal applied to $V_{in}$ [see Fig.~\ref{fig:circuitdiagram}~(a)]. }
  \label{fig:bodeplot}
\end{figure}

In Fig.~\ref{fig:bodeplot}, we present the corresponding Bode plot of the clamp switch. Fig.~\ref{fig:bodeplot}~(a) shows the small-signal attenuation factor, $\eta_{S} = 20 \log_{10} (V_{out, 0}/V_{in, 0})$, where $V_{out, 0}$ and $V_{in,0 }$ are fitted amplitudes of the respective sinusoids. The plot demonstrates attenuation factors ranging from about -55~dB to -40~dB at lower frequencies, 
corresponding to factors between 100 and 600 in noise amplitude suppression. At frequencies above about 10~kHz, the attenuation levels off at about -24~dB, corresponding to a factor of about 15 in noise amplitude suppression. Fig.~\ref{fig:bodeplot}~(b) reveals a phase shift ranging from $\sim 70^\circ$ at low to $\sim - 20^\circ$ at high frequency.

Fig.~\ref{fig:bodeplot} shows the high effectiveness of the clamp switch as a noise-suppression device. It is particularly effective at frequencies below 1~kHz, which covers ubiquitous 60-Hz noise and overtones thereof. The trends in Fig.~\ref{fig:bodeplot} reflect the fact that at low frequencies the diode capacitance $C_D$ and the load resistance $R_L$ dominate the response, leading to a complex $V_{out}/V_{in} = {\rm{i}} 2 \pi f_{rt} C_D R_L$ and a (theoretical) slope of $\eta_S$ of 2 in Fig.~\ref{fig:bodeplot}~(a). Hence, the clamp switch provides excellent noise attenuation at low frequencies, where the noise spectral density usually is the largest.
At high frequencies, the load and diode capacitances, $C_L$ and $C_D$, dominate the response, leading to a fixed and real $V_{out}/V_{in} = C_D / C_L$. These limits follow from the circuit analysis given in the Appendix. The minor deviations of Fig.~\ref{fig:bodeplot}~(b) from the analysis in the Appendix are attributed to unaccounted-for stray reactances.

From the described trends we see that $R_L$ should be kept small, while observing applicable load power and HV current sourcing constraints as well as over-voltage-protection requirements on $V_{offset}$. The high-frequency attenuation improves with increasing $C_L$, at the expense of increasingly poor dynamic large-signal response (see Section.~\ref{subsec:acresponse}).
Generally, $C_L$-values on the order of the diode present a good compromise. 
For fast general-purpose diodes, $C_L \sim 100$~pF can be suitable, corresponding to about one meter of coax connection cable from $V_{out}$ to the electrode(s) in the Rydberg experiment.

\subsection{Large-signal response of the clamp switch}
\label{subsec:acresponse}

\begin{figure}[t]
 \centering
  \includegraphics[width=0.48\textwidth]{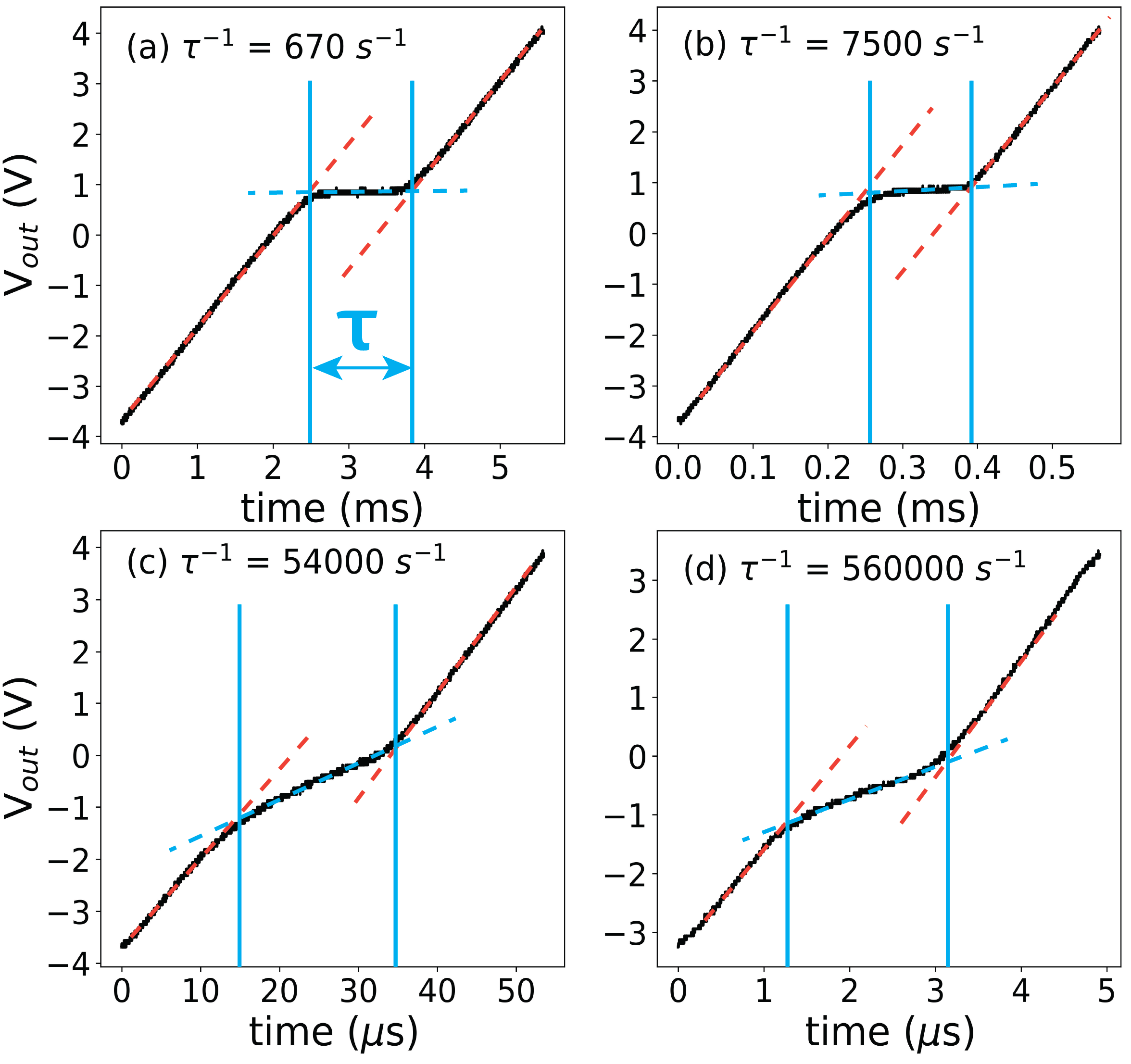}
  \caption{Response of the clamp switch to large-signal linear ramps. (a)-(d) Measured output $V_{out}$ of the clamp switch 
  for voltage ramps of increasing slew rate applied to $V_{in}$. As the slew rate increases, the clamping efficiency is reduced, and the clamping plateau shifts to earlier times by $\sim 1~\mu$s.} 
  \label{fig:acresponse}
\end{figure}

While the small-signal analysis provides useful noise suppression metrics of the circuit that are relevant to the Rydberg-atom manipulation and control phase, an analysis of the large-signal response is required to characterize the behavior during the application of the FI pulse for Rydberg-atom detection.

\begin{figure}[t]
 \centering
  \includegraphics[width=0.45\textwidth]{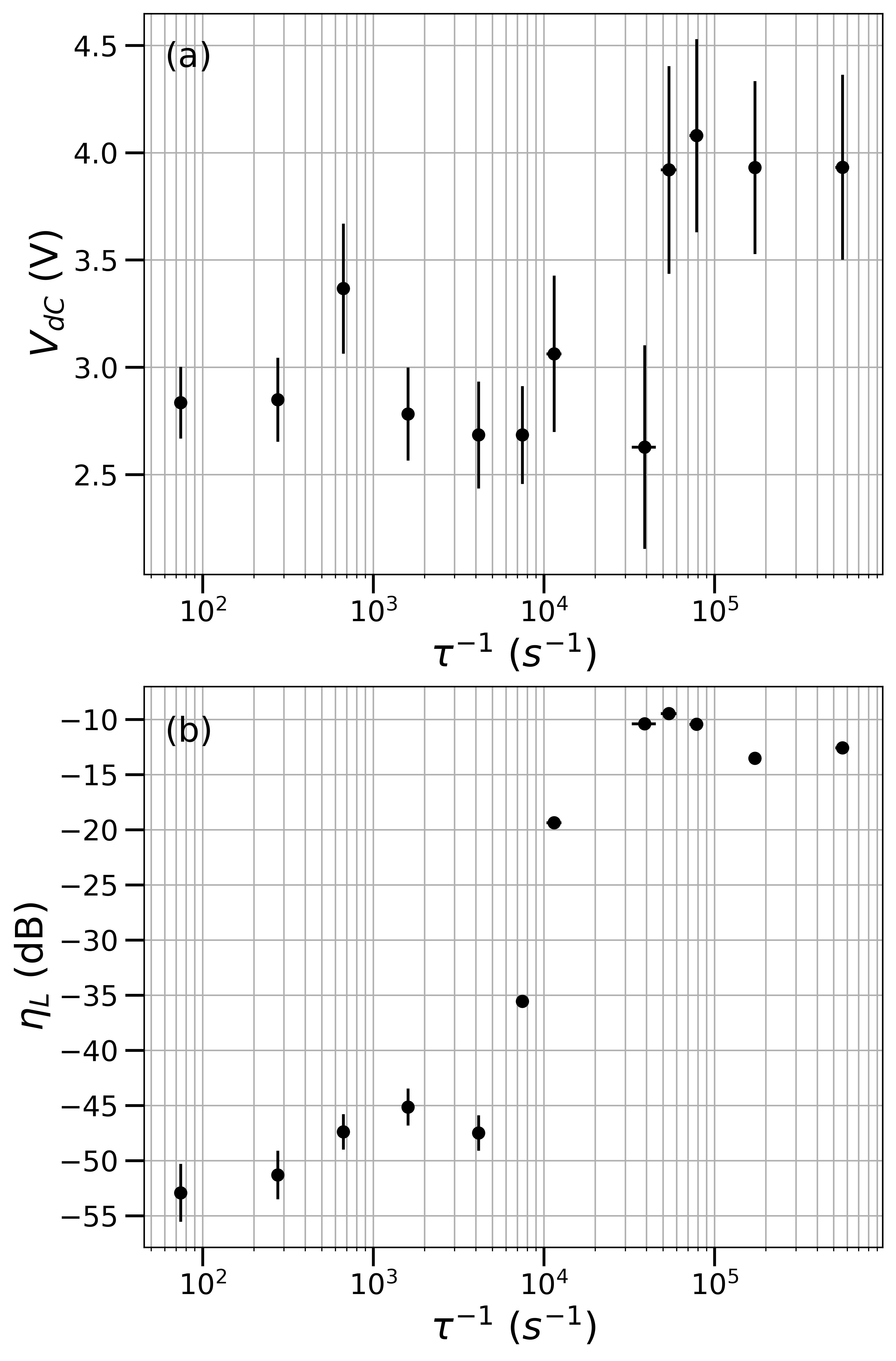}
  \caption{Measured large-signal dynamic clamping range $V_{dC}$ (a) and attenuation $\eta_L$ versus $\tau^{-1}$ (b). See text for details.}
  \label{fig:bigsignal response}
\end{figure}

Our large testing signal consists of a triangular waveform of a varied slew rate applied to $V_{in}$. Specifically, we apply symmetric saw-tooth signals with a peak-to-peak amplitude of 10~V and frequency $f_{ac}$ to $V_{in}$, corresponding to slew rates of $dV_{in}/dt  = f_{ac} \times 20$~V. The offset {input of the clamp switch [see Fig.~\ref{fig:circuitdiagram}~(a)]} is held at $V_{offset} = 1$~V. In Fig.~\ref{fig:acresponse}, we display the response $V_{out}(t)$ during the rising sections of $V_{in}(t)$ for $f_{ac}$ varied from 0.1~kHz to 100~kHz in steps of factors of 10. As $V_{in}$ passes through 1~V (the offset voltage), we observe the clamping effect, which amounts to a relatively flat plateau in $V_{out}$ that extends over some time [similar to Fig.~\ref{fig:circuitdiagram}~(b)]. 
For an empirical determination of the clamping time, we first draw two lines through the fast-rising, approximately linear segments of $V_{out}(t)$ before and after the clamping phase (dashed red lines in Fig.~\ref{fig:acresponse}). Then we approximate $V_{out}(t)$ during the relatively flat clamping plateau by the third line (dashed blue lines in Fig.~\ref{fig:acresponse}). 
The lines intersect before and after the clamping plateau, as indicated in Fig.~\ref{fig:acresponse} by solid vertical blue lines. The time difference between the intersection points defines the clamping time $\tau$. Then the dynamic clamping range $V_{dC}$ is defined as
 \begin{equation}
     V_{dC} = \frac{d V_{in}}{dt} \tau \quad.
 \end{equation}
The effectiveness of the clamp is characterized by the slope, $k$, of the lines through the clamping regions in Fig.~\ref{fig:acresponse}. As an attenuation metric similar to the one used in Fig.~\ref{fig:bodeplot}~(a), we define the large-signal attenuation factor (in dB) as
\begin{equation}
    \eta_{L} = 20 \log_{10} \left(\frac{k}{d V_{in}/dt}\right) \quad .
\end{equation}

In Fig.~\ref{fig:bigsignal response}~(a) and~(b), we show the dynamic clamping range $V_{dC}$ and attenuation factor $\eta_{L}$ versus $\tau^{-1}$, respectively. The value of $V_{dC}$ increases by $\sim 25\%$, with most of the increase occurring around $\tau^{-1} \sim 50 \times 10^3$~s$^{-1}$. We attribute the increase in $V_{dC}$ to a moderate shift of the intersection points in Fig.~\ref{fig:acresponse} away from the plateau center. The large-signal attenuation $\eta_L$ diminishes from about -55~dB at low $\tau^{-1}$ to about -15~dB at large $\tau^{-1}$, with most of the change occurring around $\tau^{-1} \sim 10^4$~s$^{-1}$. 

These observations are largely in line with the frequency dependence of the small-signal response found in Sec.~\ref{subsec:small} and are explained with the circuit model presented in the Appendix. The model generally confirms that the reduction in $\eta_L$ at high slew rates is due to capacitive time constants associated with $C_L$ and $C_D$. The detailed time dependence of $V_{out}(t)$ at high frequencies is fairly complex, including a shift of the clamping plateau to earlier times and a variation of the shape of the plateau at high slew rates. These details arise from the nonlinearity of $R_D$ and $C_D$ and are reproduced by the model in the Appendix as well.

We wish to emphasize that the large-signal response is decoupled from the intended use of the clamp switch as a small-signal noise attenuator during Rydberg-atom manipulation and control, where $V_{in}$ and $V_{offset}$ are varied in sync to keep their difference well within the clamping range $|V_{in} - V_{offset}| \lesssim V_C/2$. The large-signal response matters primarily for the exact timing of the FI, because the clamp switch may delay the FI pulse by a few $\mu$s (see Appendix). In practice, this does not present a problem, because atom counting gates for Rydberg atom detection are empirically adjusted to the FI-ramp signal applied to $V_{in}(t)$. The empirical procedure allows us to account for FI delays caused by the clamp switch.

\section{Application in Rydberg atom spectroscopy}
\label{sec:appl}

\subsection{Electrode control scheme}
\label{subsec:electrode}

\begin{figure}[t]
 \centering
  \includegraphics[width=0.48\textwidth]{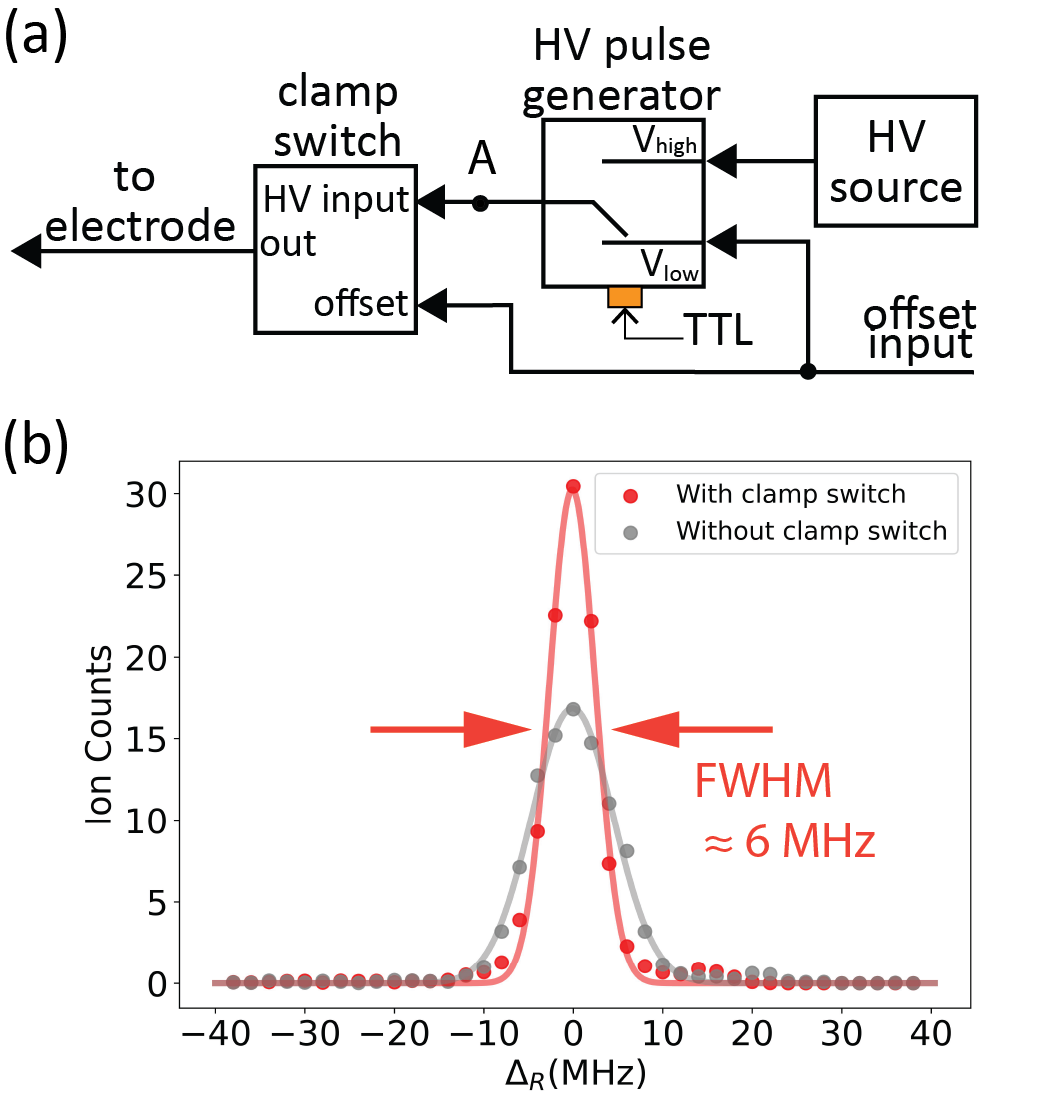}
  \caption{Sample application of the clamp switch in Rydberg-atom spectroscopy. (a) Circuit diagram for controlling the FI electrode in the utilized experimental setup. (b) An example of a Rb  $50F_{5/2}$ Rydberg spectrum obtained without ({grey dots - experimental data, gray line - Gaussian fit}) and with clamp switch ({red dots - experimental data, red line - Gaussian fit}).}
  \label{fig:spectrum}
\end{figure}

As mentioned throughout this paper, the clamp switch is useful for applications in Rydberg-atom physics. An example of its implementation is shown in Fig.~\ref{fig:spectrum}~(a). There, a fixed HV source and a variable signal $V_{offset}$ are connected to a pair of inputs of a fast, TTL-controlled HV switch (HV pulse generator). The output of the HV switch is connected to the HV input, $V_{in}$, of the clamp switch. A low-noise copy of $V_{offset}$ is routed to the offset input of the clamp switch. The clamp switch output, $V_{out}$, is toggled between the low-noise signal $V_{offset}$ for high-fidelity Rydberg-atom manipulation and the HV signal for readout via FI. The output $V_{out}$ is connected to the relevant electrode(s) in the experimental setup. The value of $V_{offset}$ applied to both the HV switch and the offset input of the clamp switch is adjustable and can be scanned or varied as a function of time. In this way, during the Rydberg-atom manipulation and control phase of the experiment, the low-noise control voltage $V_{offset}(t)$ is passed directly to the setup, while its higher-noise copy present at the output of the HV switch is connected to the offset input of the clamp switch, guaranteeing the validity of the clamping condition $|V_{in} - V_{offset}| < V_C/2$. {Complex Rydberg-atom manipulation and control sequences are possible via dynamical control of $V_{offset}(t)$. }

\subsection{Spectroscopic estimation of stray electric field}
\label{subsec:spectrum}

In Fig.~\ref{fig:spectrum}~(b), a typical Rydberg spectrum obtained when using the clamp switch is shown. The details of the experimental setup are described in~\cite{Chen2014praatomtrapping, ionsourcepaper}. Rubidium (Rb) atoms are laser-cooled in a magneto-optical trap (MOT) from which they are loaded into an optical lattice. After turning off the lattice, the atoms undergo the three-photon laser excitation sequence $\ket{5S_{1/2}, F=3} \rightarrow \ket{5P_{1/2}, F=3} \rightarrow \ket{5D_{3/2}, F=4} \rightarrow \ket{50F_{5/2}}$. The {lower-stage} lasers (795~nm and 762~nm) are locked to the transitions $\ket{5S_{1/2}, F=3} \rightarrow \ket{5P_{1/2}, F=3}$ and $\ket{5P_{1/2}, F=3} \rightarrow \ket{5D_{3/2}, F=4}$, respectively. The third laser at 1256~nm is {frequency-}scanned (detuning $\Delta_R$) and {drives the $\ket{5D_{3/2}, F=4} \rightarrow \ket{50F_{5/2}}$ transition}. The dc electric field is fine-adjusted to near zero using several low-voltage and low-noise field-compensation electrodes. After laser excitation, the voltages on two of the electrodes are ramped up to HV through the clamp switch, as shown in Fig.~\ref{fig:spectrum}~(a), to field-ionize the Rydberg atoms. Ion counts detected with a micro-channel plate detector~\cite{JAGUTZKI2002244} are recorded with a computer-controlled data acquisition system. At test point A in Fig.~\ref{fig:spectrum}~(a), the voltage signal carries noise with a $V_{RMS}$ of 11.4~mV. 

{In Fig.~\ref{fig:spectrum}~(b), we show two experimental spectra taken with and without the clamp switch (dots) and the corresponding Gaussian fits (solid lines). Each spectrum is an average of seven scans.} The Rb $50F_{5/2}$ atoms are highly sensitive to stray electric fields due to their large principal ($n=50$) and angular momentum ($\ell=3$) quantum numbers~\cite{hanpra2006, ionsourcepaper}. The full width at half maximum (FWHM) of the Rydberg line shown in Fig.~\ref{fig:spectrum}~(b) is 10.7(2)~MHz without the clamp switch. When the clamp switch is used, the FWHM is reduced to 6.0(1)~MHz. Moreover, the clamp switch increases the maximum ion count by nearly a factor of two.

The contribution of radiative and thermal decay at 300~K to 
the $50F_{5/2}$ linewidth is about 2~kHz and is negligible. The combined linewidth of the excitation lasers is estimated to be~$\lesssim$~2~MHz~\cite{cardman2021,ionsourcepaper}. The observed linewidth of 6~MHz mostly arises from stray electric fields and Rydberg atom interactions, including long-range multipolar interactions~\cite{Sassmannshausen2016, Bai2024}. Since the linewidth contributions add up in quadrature, the broadening due to stray electric fields has an upper limit of {about 5~MHz, corresponding to a standard deviation of $\sigma_f \approx 2$~MHz. The dc Stark shift in an electric field $E$ is $-\alpha E^2/2$, where $\alpha$ is the dc polarizability. Since the Stark broadening was minimized via small low-noise control voltages applied to compensation electrodes, we may assume that the electric field has a Gaussian probability distribution with an average near 0 and a standard deviation $\sigma_E$, where $\sigma_E$ describes the electric-field noise. The field-noise-induced Stark broadening $\sigma_f \approx \alpha \sigma_E^2 /\sqrt{2}$. Also, for the 50$F$ states of Rb we calculate an $m$-averaged value of $\alpha \approx 1.3 \times 10^4$~MHz/(V/cm)$^2$, where $m$ is the magnetic quantum number (see Sec.~\ref{subsec:stark} and~\cite{Reinhard2007}). The electric-field noise in Fig.~\ref{fig:spectrum}~(b) is then estimated at $\sigma_E \sim 10$~mV/cm. }

\subsection{Field calibration using Stark spectroscopy}
\label{subsec:stark}

In the next application example, we perform Stark spectroscopy of the $50F$ Rydberg state by varying the value of $V_{offset}$ on the clamp-switch circuit in Fig.~\ref{fig:spectrum}. After compensation of stray electric fields, the electric field $E$ in the region of the atoms is given by $E = \kappa V_{offset}$. Here, we determine the calibration factor $\kappa$. 

\begin{figure}[t!]
 \centering
  \includegraphics[width=0.4\textwidth]{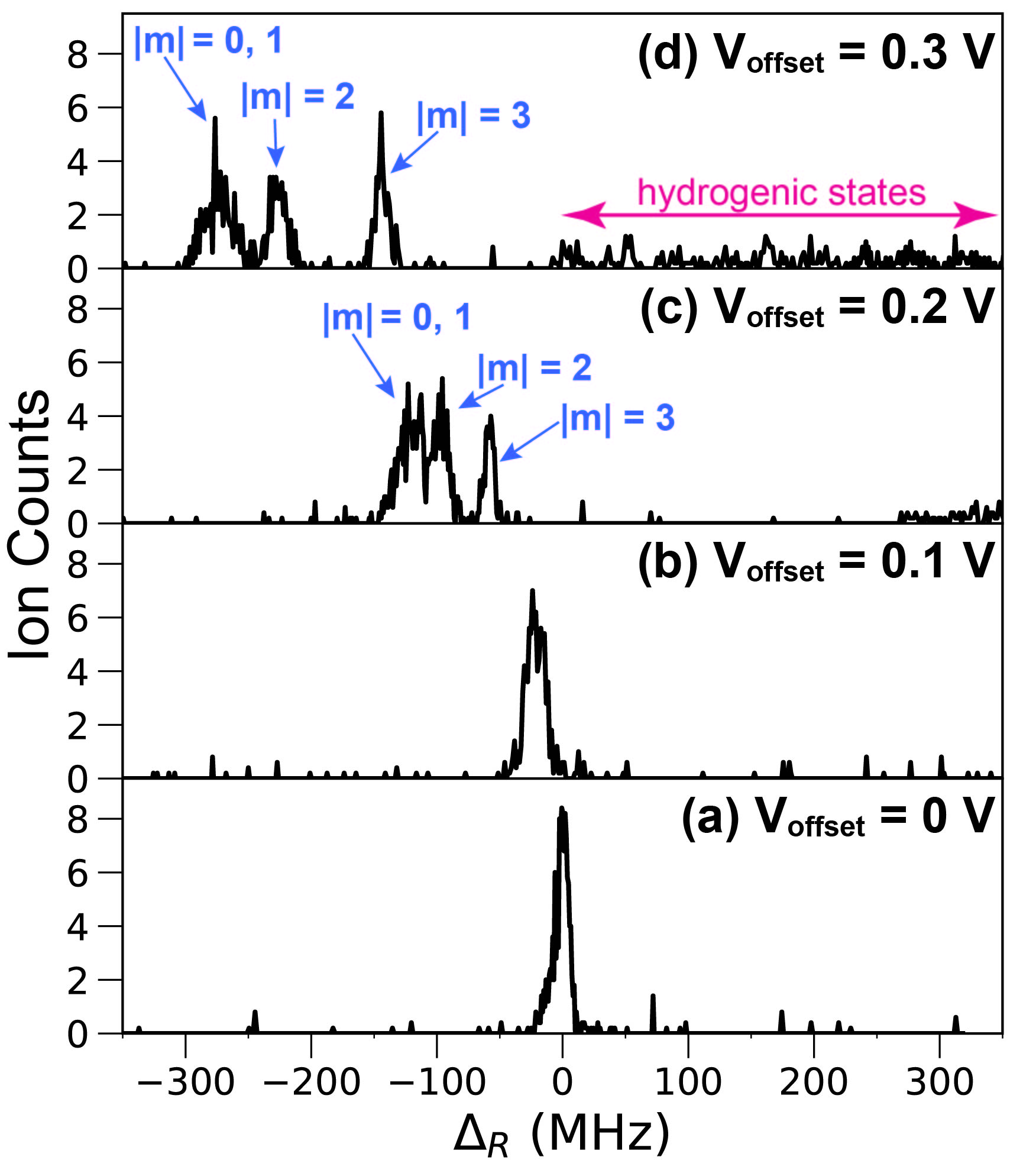}
  \caption{Stark spectroscopy of the $50F_{5/2}$ state and the neighboring hydrogenic states for the indicated voltages $V_{offset}$, which are applied to the clamp switch. }
   \label{fig:stark}
\end{figure}

Fig.~\ref{fig:stark} shows Stark spectra for the indicated values of $V_{offset}$. 
With increasing $V_{offset}$, the initial single atomic line [Fig.~\ref{fig:stark}~(a)] first broadens [Fig.~\ref{fig:stark}~(b)] and then splits into three resolved lines [Figs.~\ref{fig:stark}~(c) and (d)]. The lines correspond to the magnitudes of the allowed orbital magnetic quantum numbers, $|m| = 0, 1, 2, 3$, as the utilized Rydberg state has $\ell=3$. The states with $m$~=~0 and 1 are not resolved, and the corresponding feature in Figs.~\ref{fig:stark}~(c) and (d) is broadened due to the Stark-shift difference of the underlying unresolved states. The lines for $|m|=$ 2 and 3 are resolved, allowing us to determine the calibration factor $\kappa$.

We first note that the inverted $50F_J$ fine structure splitting of 1.3~MHz amounts to less than $0.5\%$ of the largest dc Stark shifts in Fig.~\ref{fig:stark}. Hence, the time-independent quantum states are, within a good approximation, given by $\vert n, \ell, m \rangle$, and the electron spin is irrelevant. From perturbative calculations~\cite{cardman2021} for $50F$, we obtain a set of theoretical dc polarizabilities $\alpha(m) =[17.1, 16.1, 13.1, 8.1]~\times~10^3$~MHz$/$(V/cm)$^2$ for $|m| = $~0, 1, 2, and 3 respectively. 
Alternatively, we may fit the line positions in calculated Stark maps with equations of the form $\Delta(m) = - \alpha(m) E^2 /2$ over a field range $|E| \lesssim 0.1$~V/cm. This yields a second set of theoretical polarizabilities, $\alpha(m) =[ 16.8, 15.9, 12.9, 7.84]~\times~10^3$~MHz$/$(V/cm)$^2$ for $|m| = $~0, 1, 2, and 3. The theoretical $\alpha(m)$-values obtained by the two methods differ by $\lesssim 3\%$, with the deviation attributed to higher-order Stark shifts at $|E|>0$ and residual fine-structure shifts.

The shifts of the Stark lines for $|m|=2$ and 3 measured in Figs.~\ref{fig:stark}~(c) and (d) are given by $- \alpha(m) (\kappa V_{offset})^2/2$. Using the above theoretical values for $\alpha(m)$, the calibration factor $\kappa$ is found to be $\kappa=$~0.628(9)~cm$^{-1}$. The uncertainty of $\kappa$ of $1.5\%$ follows from the sub-MHz uncertainty of the line positions and the $0.2~\times~10^3$~MHz$/$(V/cm)$^2$ variation of the theoretical $\alpha(m)$. 

We note that the spectrum at $E$~= 0~V/cm in Fig.~\ref{fig:stark}~(a) is broader than the one in Fig.~\ref{fig:spectrum}~(b) because we slightly increased the power of the 762-nm excitation laser. This results in a higher Rydberg atom number, which reduces statistical noise but enhances  line broadening and asymmetry~\cite{carroll2004,han2009,beguin2013} caused by Rydberg-atom interactions. Line shifts and broadening can also be caused {by ions produced by Penning ionization}~\cite{Barbier2, liprl2005, vincent2013} and photo-ionization~\cite{duncan2001,ionsourcepaper, cardman2021}.

Rydberg states with orbital quantum numbers $\ell>3$, often called hydrogenic states due to their quantum defects being close to 0, experience large linear Stark shifts~\cite{gallagher, zimmerman1979}. In Figs.~\ref{fig:stark}~(c) and~(d), hydrogenic Stark states give rise to the signals observed on the right, as indicated. The hydrogenic Stark states offer an alternative method for electric-field calibration and can be utilized to assess Holtsmark fields caused by ions~\cite{ionsourcepaper, demura}.

\section{Discussion and conclusion}
\label{sec:concl}

In this paper, we have described a clamp switch circuit that is used to prevent electric noise from entering control electrodes inside high-vacuum chambers.{The clamp switch is beneficial} in apparatuses for Rydberg-atom studies and applications. The switch is particularly useful for electrode(s) that serve a dual purpose as a conduit for both low-noise high-fidelity signals for experimental control, and for (typically noisy) high-voltage pulses, which are required, for example, for electric-field ionization of atoms. We have studied and explained the small-signal noise suppression characteristics as well as the large-signal behavior of the device. Appendix with the circuit analysis is provided. Speed limitations due to resistances and capacitances of the diodes and the load are discussed. We have presented two exemplary applications, one - on line narrowing via field-noise reduction achieved by the clamp switch, and the other - on electric-field calibration.

In Fig.~\ref{fig:spectrum} we have used the highly polarizable and electric-field-sensitive $50F$ Rydberg state to null the stray electric field to residual $\sigma_E \sim 10$~mV/cm. In applications in quantum information and simulation, Rydberg levels of the type $nS_{1/2}$ are often preferred because of their low electric polarizabilities and largely isotropic interactions. For example, the polarizability of the Rb state $53S_{1/2}$, which is close in atomic energy to the $50F$ state, is $\alpha = 0.76 \times 10^2$~MHz/(V/cm)$^2$. This is several orders of magnitude lower than the polarizability of $50F$, which is $\alpha \sim 1.3 \times 10^4$~MHz/(V/cm)$^2$ (see Sec.~\ref{subsec:spectrum}). An electric-field noise of $\sigma_E \sim 10$~mV/cm translates into a decoherence rate on the order of $\sim 10$~kHz for $53S_{1/2}$. This case study exemplifies that field nulling should be performed with highly polarizable states, to then conduct science with less polarizable states. We have found the clamp switch to be invaluable in these steps.

Here we have used 1N4007 diodes, which are fast, general-purpose diodes. In setups with low load capacitance, these may be replaced with faster diodes, such as 1N4148. Our circuit model in Appendix shows that users will benefit from carefully selecting diodes for their specific timing sequences and load conditions. We have found that a combination of diode parameters from data sheets, SPICE models~\cite{spicebook}, and user-performed tests yields the best match between model results and clamp-switch test data. 
 
The clamp switch has applications in a wide range of Rydberg-atom experiments that utilize detection via (state-selective) field ionization. These include fundamental-physics research, such as measurements of the Rydberg constant~\cite{Hare1993, Jentschura2008, Ramos2017} and dark-matter searches~\cite{haseyama2008, SPIE_GR, graham2024}, and a plethora of applications in quantum information science and sensing (see Sec.~\ref{sec:intro}). The application samples presented here as well as {the} Appendix will assist users with adapting the details of the circuit design to a range of conditions.

\section*{Acknowledgments}
\label{sec:acknowledgments}

The work was supported by NSF Grant No. PHY-2110049. A.D. acknowledges support from the Rackham Predoctoral Fellowship at the University of Michigan. I.H. was supported by the University of Michigan 2024 REU program, funded by NSF Award No. PHY-2149884. We also thank previous group members for their early contributions to this work.

\section*{Appendix: Circuit modeling}\label{section:app}

\begin{figure}[t!]
 \centering
  \includegraphics[width=0.47
\textwidth]{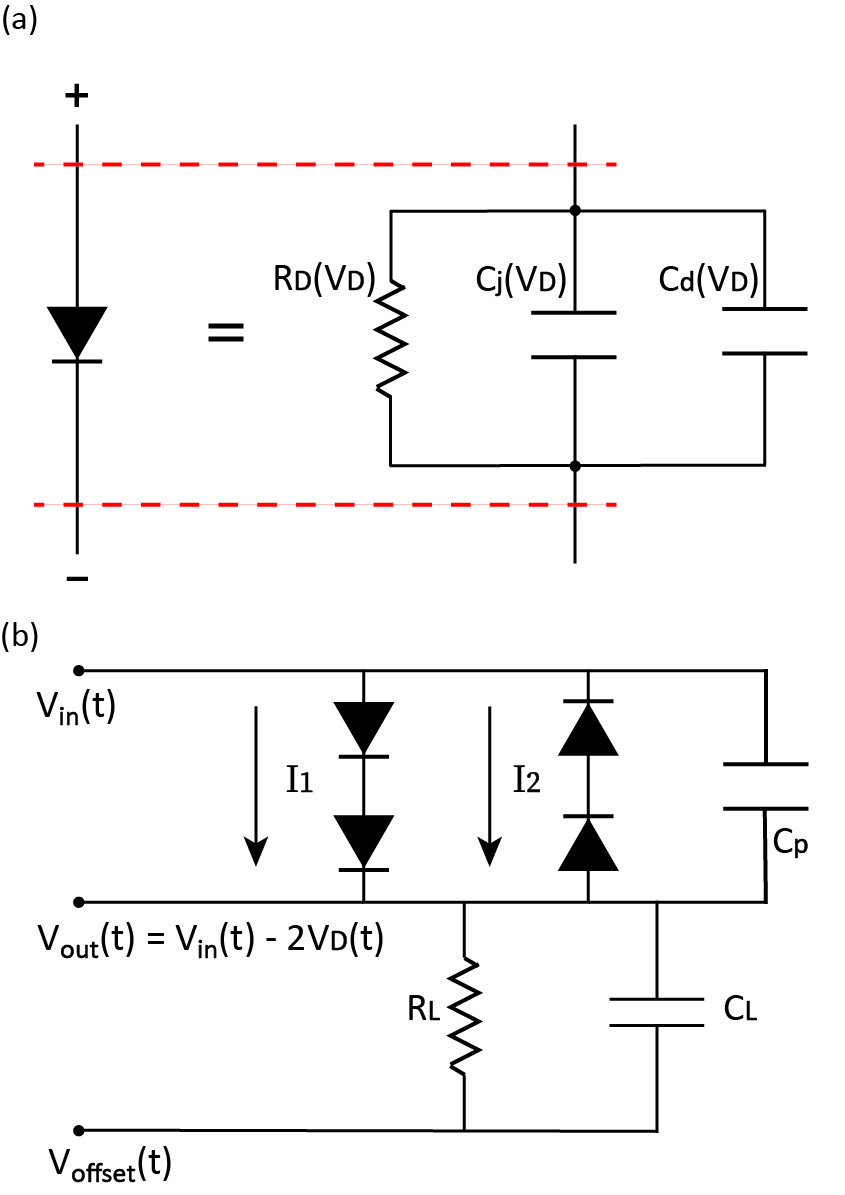}
  \caption{Circuit model with $K=2$ diodes per leg. (a) The single-diode resistance $R_D$ and junction and diffusion capacitances~\cite{Lucia_1993}, $C_j$ and $C_d$, depend on the single-diode voltage, $V_D$. Positive $V_D$ correspond with forward bias. (b) Circuit model with fixed load resistance, $R_L$, and capacitance, $C_L$. The tuning capacitor $C_P$ across the (entire) diode block is optional. 
} 
  \label{fig: diode model}
\end{figure}

The circuit model is illustrated in Fig. \ref{fig: diode model}.
Since in the clamp switch the range of the single-diode voltage, $V_D$, is limited to values above the reverse breakdown voltage, the resistive component of a single diode,
\begin{equation}
    R_D = \frac{V_D}{I_D} \quad, 
\end{equation}
follows from the current-voltage relation~\cite{shockley,pnhighfrehighvolta,smsze}    

\begin{equation}
    I_D = I_s [\exp \left(\frac{V_D}{N \cdot V_t}\right) - 1] \quad.
    \label{eq: diode iv}
\end{equation}

Here $I_s$ is the saturation current and $N$ is the emission coefficient. The thermal voltage $V_t = \frac{k_B T}{e}$ is expressed in terms of the Boltzmann constant $k_B$, the temperature $T$, and the elementary charge $e>0$.  

The diode capacitance, $C_D$, is the sum of junction and diffusion capacitances. The junction capacitance, $C_j$, is mainly due to immobile charges in the depletion region. It is dominant for $V_D \lesssim 0.35$~V and decreases moderately with decreasing $V_D$, {\sl{i.e.}} with increasing reverse bias and increasing width of the depletion region. For $V_D \gtrsim 0.35$~V, the exponentially increasing diffusion capacitance, $C_d$, rapidly dominates.

The diode capacitances are highly non-linear. For $V_{D} < FC \cdot V_j$, the junction capacitance 
\begin{equation}
    C_j = C_{j0} \left(1 - \frac{V_D}{V_j}\right)^{-M} \quad,
    \label{eq:cj}
\end{equation}
and for $V_D \geq FC \cdot V_j$
\begin{equation}
    C_{j} = \frac{C_{j0}}{(1-FC)^{M+1}} \cdot \left(1-FC \cdot (M+1) + \frac{M\cdot V_D}{V_j}\right) \quad.
\end{equation}
Here, $C_{j0}$ is the zero-bias junction capacitance and $V_j$ is the contact potential. The parameter $M$ is a grading exponent that is used to change the slope of the curve $C_j(V_D)$. The parameter $FC$ is used to model $C_{j0}$ under forward bias condition and to ensure a smooth and numerically stable transition in modeling junction capacitance from reverse-biased or weakly forward-biased conditions to stronger forward-biased conditions.

The diffusion capacitance, $C_d$, arises from an excess of minority carriers near the edges of the depletion region, which scales with the diode current. Therefore, $C_d$ is often neglected in reverse bias, rises exponentially near zero bias in sync with the forward current, and becomes dominant for the forward bias $V_D \gtrsim 0.35$~V. When the diode is switched from forward to reverse, a current will flow for a short period of time in the negative direction until the minority charge is removed. This period is called the diode transit time $TT$. For $V_D >= -5 \cdot N \cdot V_t$, $C_d$ can be expressed as
\begin{equation}
    C_d = \frac{TT \cdot I_s}{N \cdot V_t} \exp{\frac{V_D}{N V_t}} \quad ,
\end{equation}
while for $V_D < -5 \cdot N \cdot V_t$ we set

\begin{equation}
 C_d = \frac{TT \cdot I_s}{N \cdot V_t} \exp(-5) \quad.
 \label{eq:cd}
\end{equation}
SPICE models provide typical parameters for specific diodes~\cite{spicebook}. Since in the clamp switch $V_D$ is always well above the breakdown voltage, diode behavior near and {beyond} the breakdown is irrelevant here.

\begin{figure}[t!]
\centering
\includegraphics[width=0.45\textwidth]{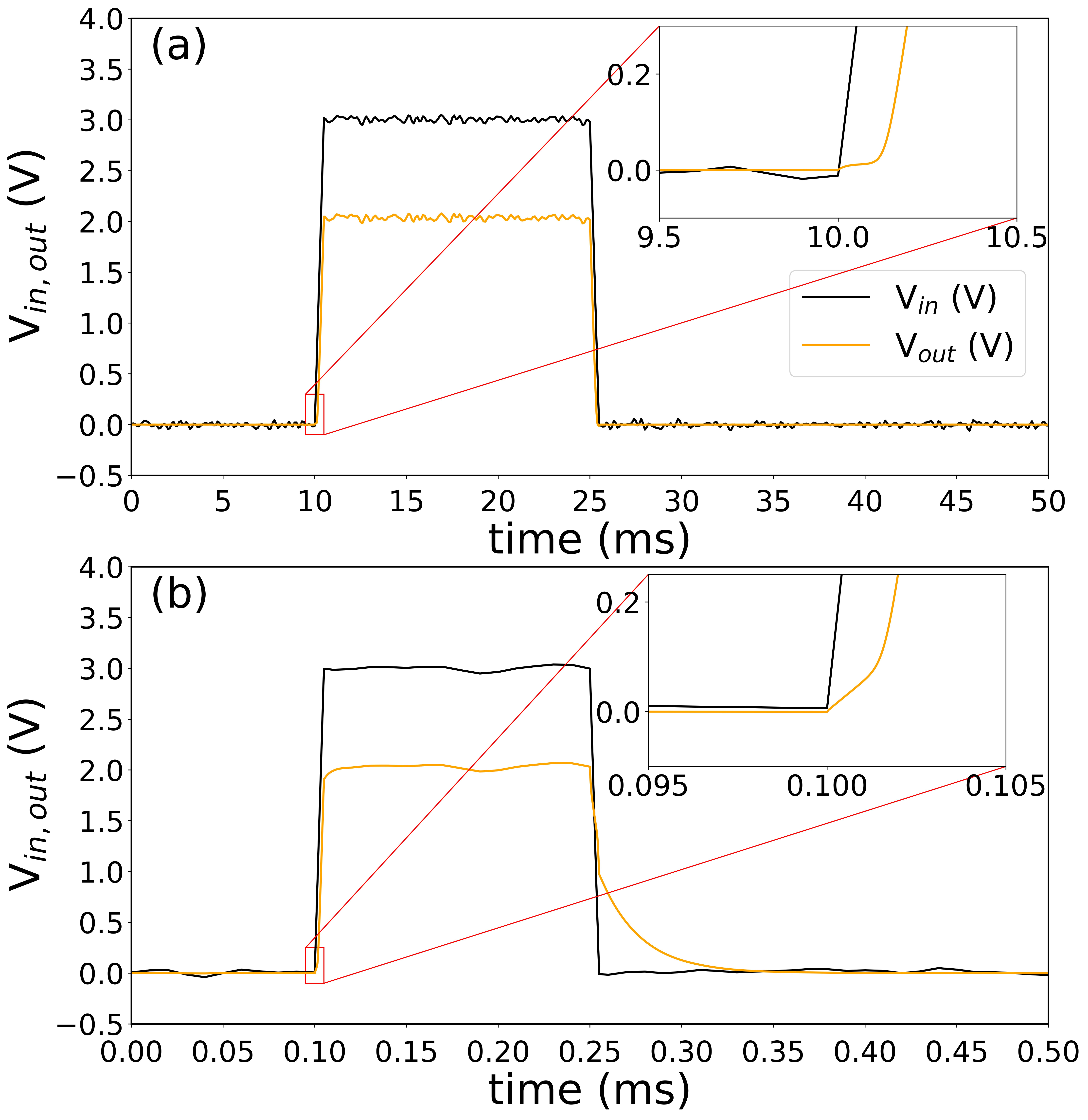}
\caption{{Simulated single-pulse response of the clamp switch. We display simulated 
$V_{out}(t)$ output (yellow) and input signals $V_{in}(t)$ (black) for a slow- (a) and a fast-pulse (b) case. The input pulse signal is superimposed with random noise with the mean value of zero and RMS value of 26~mV. The inset shows a magnified view at the pulse onset.}}
  \label{fig:pslope response}
\end{figure}

Based on the circuit in Fig.~\ref{fig: diode model}, the current through the left ($I_1$) and the right ($I_2$) leg of diodes and the tuning capacitor $C_P$ is
\begin{align}
   \nonumber I &= I_1 + I_2 + I_P\\ \nonumber
    \nonumber &= I_{1C} + I_{1R} + I_{2C} + I_{2R} + I_P \\
    &= \frac{dV_D}{dt} 
    \big[C_1 +  V_D C'_1 + C_2 - V_D C'_2 + K C_P \big] + I_{1R} + I_{2R} 
    \label{eq:diodepairI}
\end{align}
where $C_{1}=C_D(V_D)$, and $C_{2}=C_D(-V_D)$, and  $C_D=C_j + C_d$ with Eqns.~\ref{eq:cj}-\ref{eq:cd} above, and $C'_1 = \frac{dC_D}{dV_D} (V_D)$, and $C'_2 = \frac{dC_D}{dV_D} (-V_D)$, and $I_{1R}=I_D(V_D)$ with $I_D$ from Eq.~\ref{eq: diode iv}, and $I_{2R}=I_D(-V_D)$, $K$ is the number of diodes per leg (which is 2 in our examples), and $C_P$ is the tuning capacitance. The total current $I$ also is
\begin{align}
    \label{eq:currentkic}
     I &=I_{R_L} + I_{C_L} = \frac{V_{out}}{R} + C_L \frac{dV_{out}}{dt} \quad ,
\end{align}
where $R_L$ is the load resistor connected to the offset voltage $V_{offset}$ and $C_L$ is the load capacitance. Further, $V_{out}(t)$ is given by
\begin{align}
    V_{out}(t) =V_{in}(t) - K V_D(t)
    \label{eq:diodeKL} \quad ,
\end{align}

\noindent where $V_{in}$ is the input voltage. Combining Eqns.~\ref{eq:cd}-\ref{eq:diodeKL}, 
the following non-linear first-order ordinary 
differential equation for the large-signal time-dependent behavior of the circuit  is obtained:
\begin{widetext}
\begin{equation}
    \frac{dV_D}{dt} = \frac{ \frac{1}{R_L} \big[ V_{in}(t) - K V_D - V_{offset}(t) \big]
    + {C_L} \big[ \frac{d V_{in}}{dt} (t) - \frac{d V_{offset}}{dt} (t) \big] - I_{D}(V_D) - I_{D}(-V_D)}
      {C_D(V_D) +  V_D C_D' (V_D) + 
       C_D(-V_D) -  V_D C_D'(-V_D) + K C_L + K C_P} \quad.
     \label{eq:diodepairODE}
\end{equation}
\end{widetext}

\noindent Here, the input function $V_{in}$ depends on time $t$, and the diode parameters $I_D$, $C_D$ and $C'_D$ {depend} on $V_D(t)$. In our examples, the offset voltage $V_{offset}$ is fixed, but Eq.~\ref{eq:diodepairODE} also covers the case of time-dependent  $V_{offset}$. Eq.~\ref{eq:diodepairODE} is numerically solved using, for instance, the Runge-Kutta method. We use the initial condition $V_D(t=0) =0$; any transients damp out on a time scale of tens of $\mu$s or less. 

\begin{figure}[t!]
\centering
\includegraphics[width=0.45\textwidth]{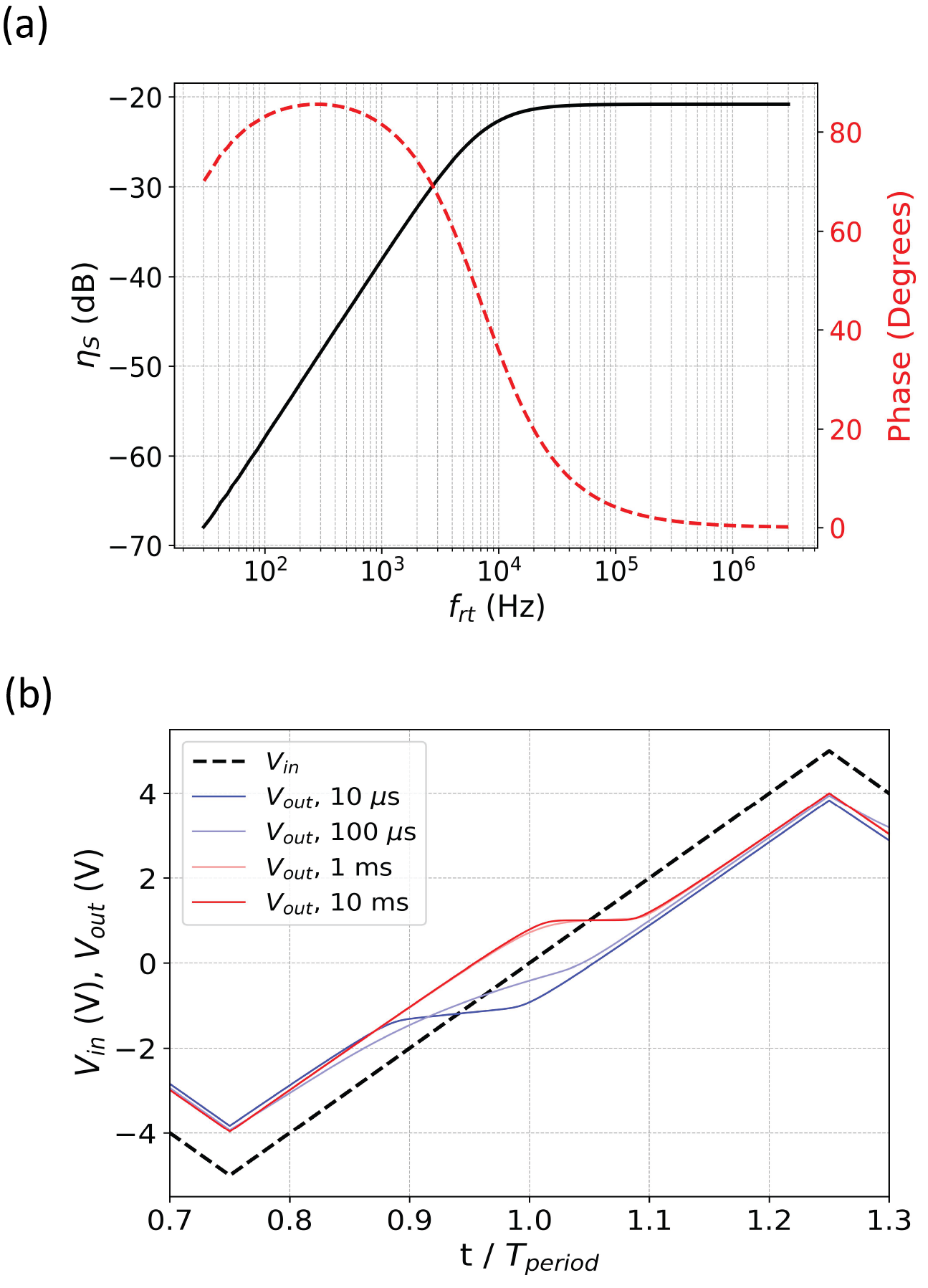}
\caption{{(a) Simulated Bode plot showing attenuation $\eta_S$ and output-signal phase versus frequency. (b) Simulated output ($V_{out}$; red curves) for large-signal triangle waveforms ($V_{in}$; black curve) for the indicated input-signal periods $T_{period}$ and an offset $V_{offset} = 1$~V. Time is shown in units of $T_{period}$.
}}
  \label{fig:simbodeplot}
\end{figure}

Eq.~\ref{eq:diodepairODE} is highly non-linear through the dependencies of the diode parameters on $V_D(t)$. We find that time steps between 1/100   and 1/1000 of the transit time, $TT$, are sufficiently fine for convergence, even for large-signal ramps on $V_{in}(t)$. For single-pulse inputs $V_{in}(t)$, which are used to model FI ramps applied to atoms, we solve Eq.~\ref{eq:diodepairODE} and plot $V_{out}(t)$ over the duration of the pulse. In Fig.~\ref{fig:pslope response} we simulate the effect of the clamp switch on a single unipolar pulse with the 26-mV RMS noise. Since we found the effect of $C_D'(-V_D)$ and $C_D'(V_D)$ {to be minor, in the present simulation we  have ignored these two terms.} For the slow pulse in Fig.~\ref{fig:pslope response}~(a), we applied the bandwidth-limited noise ranging from 10 Hz to 3 kHz, while for the fast pulse in Fig.~\ref{fig:pslope response}~(b), we applied the bandwidth-limited noise ranging from 10 Hz to 30 kHz. The slow-pulse case demonstrates the quasi-static behavior of the clamp switch for pulses with low slew rates. The fast-pulse case is suitable for FI of Rydberg atoms. Before the onset of the pulse, the RMS noise on the input is largely blocked from the output by an attenuation factor $\eta_S \sim -33$~dB and $\eta_S \sim -22$~dB for Figs.~\ref{fig:pslope response}~(a) and~(b), respectively. These simulations align well with the small-signal response in Fig. \ref{fig:bodeplot}~(a). The clamp switch delays the output pulse in (b) by about 2~$\mu$s, which would be acceptable in typical Rydberg physics experiments. The clamp switch further softens the pulse onset, a beneficial feature that helps suppress nonadiabatic Rydberg-atom dynamics at the beginning of the FI ramp. On the falling edge, $V_{out}$ has a positive tail with a decay time of about $R_L C_L$. The tail occurs because $C_L \gg C_D$ in the simulated case. The tail is inconsequential in FI because the detection sequence is complete at the time when the pulse falls.

In analogy with the circuit tests in Sec.~\ref{subsec:small} of the main text, we also model the clamp switch for small-signal harmonic input. In Fig.~\ref{fig:simbodeplot}~(a), we show a Bode plot for $V_{in}= 25$~mV$\times \sin(2 \pi f_{rt} t)$. We first compute $V_{out}(t)$ over 100 periods of the input signal and then obtain the Fourier transform of $V_{out}(t)$ at integer multiples of $f_{rt}$. The Fourier amplitudes for the fundamental $n=1$ yield the $\eta_S$ and the phases shown in Fig.~\ref{fig:simbodeplot}~(a).

For large-signal periodic input $V_{in}(t)$, we display $V_{out}(t)$ after {a one-period wait time}, which suffices for any transients to damp out. In Fig.~\ref{fig:simbodeplot}~(b) we show $V_{in}(t)$ and $V_{out}(t)$ on the rising slope of a selection of large-signal triangle waveforms, in analogy with Sec.~\ref{subsec:acresponse} in the main text. Panels (a) and (b) of Fig.~\ref{fig:simbodeplot} are in good agreement with Figs.~\ref{fig:bodeplot} and~\ref{fig:acresponse} in the main text, {with the exception of a $-20^\circ$ overall shift of the measured relative to the calculated phases. The {phase-shift deviation} is attributed to stray reactances in the testing setup.}

The above analysis and the sample data show 
that the simulations can help with diode selection, the number of diodes per leg, $K$, as well as with adjustments of $C_L$, $C_P$, and $R_L$ to satisfy user-specified requirements. The diode parameters necessary in Eqns.~\ref{eq: diode iv}-\ref{eq:cd} can be drawn from SPICE models~\cite{spicebook} and {may} have to be fine-tuned to match experimental circuit tests.

\bibliography{bibliography.bib}

\end{document}